# Discovery of a resolved white dwarf–brown dwarf binary with a small projected separation: SDSS J222551.65+001637.7AB

Jenni R. French ,[1★] Sarah L. Casewell,[1] Trent J. Dupuy,[2] John H. Debes,[3] Elena Manjavacas,[3] Emily C. Martin[4] and Siyi Xu(许偲艺) [5]

[1]*School of Physics and Astronomy, University of Leicester, University Road, Leicester, LE1 7RH, United Kingdom*
[2]*Royal Observatory Edinburgh, Blackford Hill, Edinburgh, EH9 3HJ, United Kingdom*
[3]*AURA for the European Space Agency (ESA), Space Telescope Science Institute, 3700 San Martin Drive, Baltimore, Maryland 21218, USA*
[4]*Department of Astronomy & Astrophysics, University of California Santa Cruz, 1156 High Street, Santa Cruz, California 95064, USA*
[5]*Gemini Observatory/NSF's NOIRLab, 670 N. A'ohoku Place, Hilo, Hawaii, 96720, USA*

## ABSTRACT

We present the confirmation of SDSS J222551.65+001637.7AB as a closely separated, resolved, white dwarf–brown dwarf binary. We have obtained spectroscopy from GNIRS and seeing-limited $K_s$-band imaging from NIRI on Gemini North. The target is spatially resolved into its constituent components: a 10926 ± 246 K white dwarf, with log $g = 8.214 \pm 0.168$ and a mass of $0.66^{+0.11}_{-0.06}\,\mathrm{M}_{\odot}$, and an L4 brown dwarf companion, which are separated by 0.9498 ± 0.0022 arcsec. We derive the fundamental properties of the companion from the Sonora–Bobcat evolutionary models, finding a mass of 25–53 $M_{\mathrm{Jup}}$ and a radius of 0.101–0.128 $\mathrm{R}_{\odot}$ for the brown dwarf, at a confidence level of $1\sigma$. We use WDWARFDATE to determine the age of the binary as $1.97^{+4.41}_{-0.76}$ Gyr. A kinematic analysis shows that this binary is likely a member of the thick disc. The distance to the binary is $218^{+14}_{-13}$ pc, and hence the projected separation of the binary is $207^{+13}_{-12}$ au. Whilst the white dwarf progenitor was on the main sequence the binary separation would have been 69 ± 5 au. SDSS J222551.65+001637.7AB is the third closest spatially resolved white dwarf–brown dwarf binary after GD 165AB and PHL 5038AB.

**Key words:** brown dwarfs – white dwarfs – binaries.

## 1 INTRODUCTION

The prolific discoveries of field brown dwarfs have been enabled by infrared (IR) and optical surveys such as the Two Micron All Sky Survey (2MASS), UKIRT Infrared Deep Sky Survey (UKIDSS), and Wide-field Infrared Survey Explorer (WISE) (e.g. Kirkpatrick et al. 1998; Pinfield et al. 2008; Aberasturi, Solano & Martín 2011). Despite over a thousand known field brown dwarfs, it is often challenging to determine their masses, ages, and luminosities as these parameters are degenerate due to the lack of fusion in the cores of brown dwarfs (Burrows & Liebert 1993; Auddy, Basu & Valluri 2016). Consequently, estimating brown dwarf parameters relies on models which, despite recent improvements, are sensitive to atmospheric processes that remain poorly understood, such as non-equilibrium chemistry (Tremblin et al. 2015) and the formation and presence of atmospheric clouds (Morley et al. 2014), and any resultant variability.

'Benchmark' brown dwarfs, those for which physical properties can be independently determined, can test and improve the current evolutidaonary and atmospheric models for brown dwarfs. Double-lined eclipsing brown dwarfs are ideal benchmark objects, but they are extremely rare, with only two known to date, 2MASS J05352184-0546085 (Stassun, Mathieu & Valenti 2006) and 2MASSW J1510478-281817 (Triaud et al. 2020). Since binary systems are formed from the same matter, companions can be assumed to have the same ages and metallicities as their primary stars (e.g. Moe & Di Stefano 2017). Therefore, brown dwarfs that are members of a binary system can have their physical properties estimated with evolutionary models using the age constraints of the primary (Pinfield et al. 2006), provided that the primary's age is well calibrated. These brown dwarfs can then be used as benchmarks to test and refine current brown dwarf models.

Widely separated binaries with a white dwarf primary and a brown dwarf companion are particularly valuable for identifying benchmark brown dwarfs. The white dwarf cooling age provides a lower limit on the age of the binary, which can be used to calculate the physical properties of the brown dwarf companion (Fontaine, Brassard & Bergeron 2001). Where the binary system is resolvable, meaning it is wide ($\gtrsim$ 50 au) or a common proper motion pair (ultra-wide, $\gtrsim$ 500 au, Zhang et al. 2020), it is unlikely that either the brown dwarf was affected during the evolution of the white dwarf progenitor, or that the white dwarf evolution was truncated by the presence of a brown dwarf companion (Meisner et al. 2020). Close white dwarf–brown dwarf binaries are not appropriate for identifying benchmark brown dwarfs because they undergo a period of common-envelope evolution instead of each component evolving separately (e.g. Casewell et al. 2018). Wide white dwarf–brown

★ E-mail: jf328@leicester.ac.uk

**Table 1.** Parameters of the brown dwarf companions in the confirmed wide, comoving white dwarf–brown dwarf binaries.

| System | Spectral Type | Separation (au) | Total Age (Gyr) | Reference |
|---|---|---|---|---|
| GD 165AB | L4 | 123 ± 12 | 1.2–5.5 | 1, 2, 3 |
| PHL 5038AB | L8 | 69 ± 1 | 1.9–2.7 | 4, 5 |
| LSPM 1459+0857AB | T4.5 | 16500-26500 | >4.8 | 5, 6 |
| WD 0806-661AB | Y1 | 2504 ± 4 | 1.5–2.7 | 5, 7, 8 |
| LSPM J0241+2553AB | L1 | 2380 ± 36 | <10 | 5, 9 |
| COCONUTS-1AB | T4 | 1290 ± 13 | $7.3^{+2.8}_{-1.6}$ | 5, 10 |
| LSPM J0055+5948AB | T8 | 402 ± 2 | 10 ± 3 | 5, 11 |

*Note.* References are 1: Becklin & Zuckerman (1988); 2: Kirkpatrick, Henry & Liebert (1993); 3: Kirkpatrick et al. (1999); 4: Steele et al. (2009); 5: Gaia Collaboration (2020); 6: Day-Jones et al. (2011); 7: Luhman, Burgasser & Bochanski (2011); 8: Rodriguez et al. (2011), 9: Deacon et al. (2014); 10: Zhang et al. (2020); 11: Meisner et al. (2020).

dwarf binaries are thus ideal systems to identify benchmark brown dwarfs to improve model calibrations and flag for follow-up with instruments such as NIRSpec on *JWST*.

Despite several all-sky surveys that have identified thousands of white dwarfs (e.g. Girven et al. 2011; Gentile Fusillo et al. 2019), and studies searching for ultracool companions to white dwarfs (e.g. Steele et al. 2010; Debes et al. 2011; Hogg et al. 2020), only ∼0.1–0.5 percent of white dwarfs are predicted to have a brown dwarf companion (Farihi, Becklin & Zuckerman 2005; Steele et al. 2011; Rebassa-Mansergas et al. 2019). There are currently only 7 known wide, comoving white dwarf–brown dwarf binaries, which are listed in Table 1. Their companion spectral types range from L1 to Y1 with separations ranging from 69 au to 16500–26500 au. The larger separations correspond to T-dwarf companions but this is likely a selection effect. In addition to these 7 confirmed white dwarf–brown dwarf binaries, there are also several candidates that have been identified using photometry but have not been spectroscopically confirmed (e.g. Kiwy et al. 2022: NSC J053232.31-512450.75AB, WD+L3, a=508 au; NSC J130527.23-224728.44AB, WD+L3, a=716 au), or that need more-IR data to confirm them as white dwarf–brown dwarf binaries (e.g. Meisner et al. 2020: LSR J0002+6357AB, WD+mid/late T, a=8695 au).

In this paper, we present a new white dwarf–brown dwarf binary, SDSS J222551.65+001637.7AB, which hosts an important benchmark brown dwarf. SDSS J222551.65+001637.7AB is the third closest separated resolved white dwarf–brown dwarf binary after GD 165AB and PHL 5038AB and understanding its evolution will provide insights into the formation of these rare wide white dwarf–brown dwarf binaries. Section 2 discusses this system, Section 3 describes our observations and data reduction, Section 4 presents our results, and Section 5 discusses our analysis.

## 2 SDSS J22255 1.65+001637.7AB

SDSS J222551.65+001637.7AB (henceforth SDSS J22255+0016AB) was first reported by Eisenstein et al. (2006) where SDSS J22255+0016A was identified as a hydrogen-rich DA white dwarf as part of a catalogue of spectroscopically confirmed white dwarfs from the Sloan Digital Sky Survey (SDSS). IR observations were taken as part of the UKIDSS Large Area Survey (Lawrence et al. 2007). Girven et al. (2011) identified SDSS J22255+0016AB as having a potential photometric near-IR excess, with Steele et al. (2011) suggesting the excess is indicative of a partially resolved L-dwarf companion. Fig. 1, which is reproduced from Steele et al. (2011), shows that in the SOFI *K*-band image SDSS J22255+0016AB is considerably elongated with a full width at half-maximum (FWHM) of 1.8 arcsec, suggesting a predicted separation of a < 350 au at their adopted distance of 190 ± 20 pc.

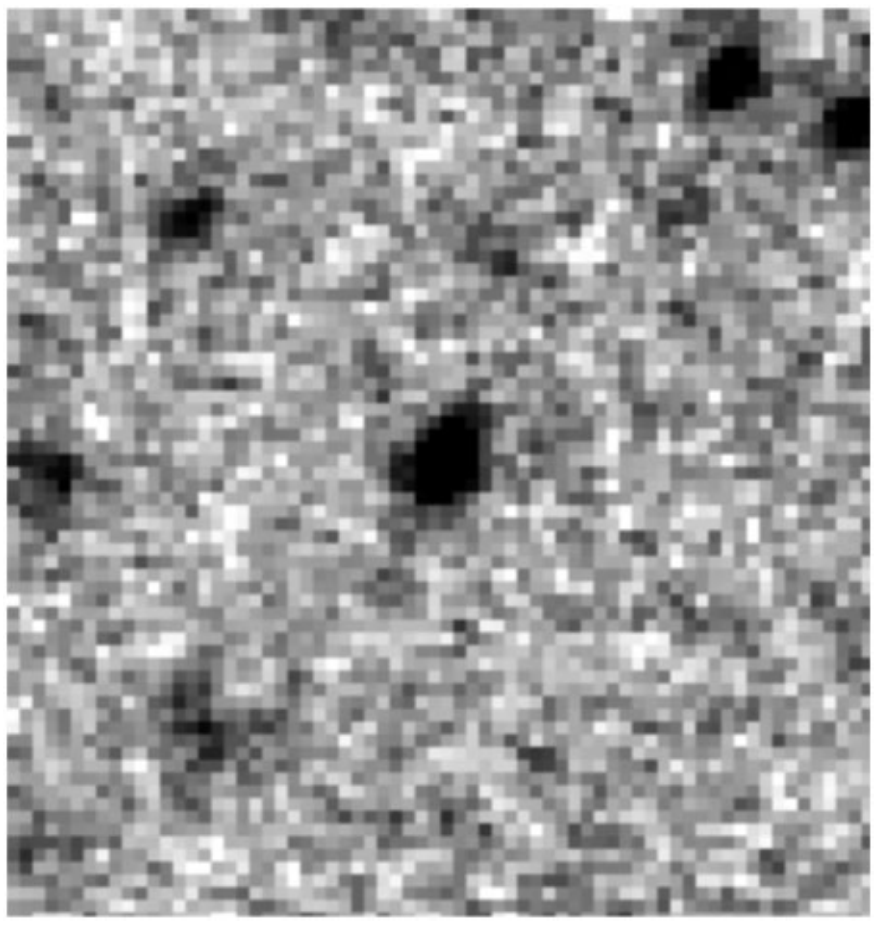

**Figure 1.** SOFI *K*-band image of SDSS J222551.65+001637.7AB showing the elongation of the central star before the components were resolvable. Reproduced from fig. 11 of Steele et al. (2011).

Eisenstein et al. (2006) fit the SDSS spectrum of SDSS J22255+0016A to a grid of model atmospheres, determining $T_{\rm eff} = 10640 \pm 94$ K and log $g = 8.16 \pm 0.09$. Anguiano et al. (2017) refined these measurements to $T_{\rm eff} = 10926 \pm 246$ K and log $g = 8.214 \pm 0.168$, with a derived distance of 226 ± 41 pc using updated model spectra and a correction for the 3D dependence of convection.

Gentile Fusillo et al. (2021) performed a photometric fit to the *Gaia* eDR3 magnitudes obtaining $T_{\rm eff} = 9370 \pm 765$ K and log $g = 7.738 \pm 0.276$. This method of fitting is less sensitive to log $g$ and relies on the *Gaia* magnitudes alone. Jiménez-Esteban et al. (2018) also performed a photometric fit of the white dwarf, and they found $T_{\rm eff} = 9000 \pm 125$ K and log $g = 6.5 \pm 0.25$ when considering the spectral energy distribution from the SDSS *u*-band to the UKIDSS *K*-band. Although these two photometric fits yield white dwarf parameters that are consistent with each other, neither of these fits account for extinction. Jiménez-Esteban et al. (2018) state that for objects at ∼200 pc, extinction should be accounted for. SDSS J22255+0016A is at a distance of $218^{+14}_{-13}$ pc, and has an extinction of $A_g = 0.323$ (Eisenstein et al. 2006), which is significant and should thus be considered.

Fig. 2 depicts a comparison of hydrogen-rich DA white dwarf models (Koester 2010) for both the derived spectroscopic and photometric parameters, alongside dereddened photometry measurements, and the SDSS spectrum of SDSS J22255+0016A. Considering the dereddened photometry, the white dwarf model that uses spectroscopic parameters ($T_{\rm eff} = 11000$ K, log $g = 8.25$) is consistent with the data, whereas the white dwarf model which uses photometric parameters ($T_{\rm eff} = 9250$ K, log $g = 7.75$) is not. This is particularly evident at shorter wavelengths, where the photometry measurements are unlikely to be influenced by the flux of the brown dwarf. The residuals in Fig. 2 for the photometric parameters are 12 percent higher than those for the spectroscopic parameters. The white dwarf model using spectroscopic parameters also better fits the depth of the hydrogen lines.

Our comparison of two Koester DA white dwarf models with the dereddened photometry measurements indicates that the spectro-scopic parameters derived by Anguiano et al. (2017) are the most appropriate. Furthermore, Groenewegen (2020) found that effective temperatures calculated from photometric fits were consistently underestimated compared to those derived from spectroscopy, with discrepancies of 2000 K in some cases. Since the Anguiano et al. (2017) parameters are derived from a spectroscopic fit, are consistent with the photometry and SDSS spectrum of SDSS J22255+0016A, and their derived distance of 226 ± 41 pc is within 1$\sigma$ of the *Gaia* eDR3 distance of $218^{+14}_{-13}$ pc, we adopt these parameters of $T_{\rm eff} = 10926 \pm 246$ K and $\log g = 8.214 \pm 0.168$ for the white dwarf. Although we start to see the contributions of the brown dwarf from the *r*-band, the brown dwarf emits mainly in the IR, and thus will not contribute significantly to the measured $T_{\rm eff}$ and $\log g$ of the white dwarf.

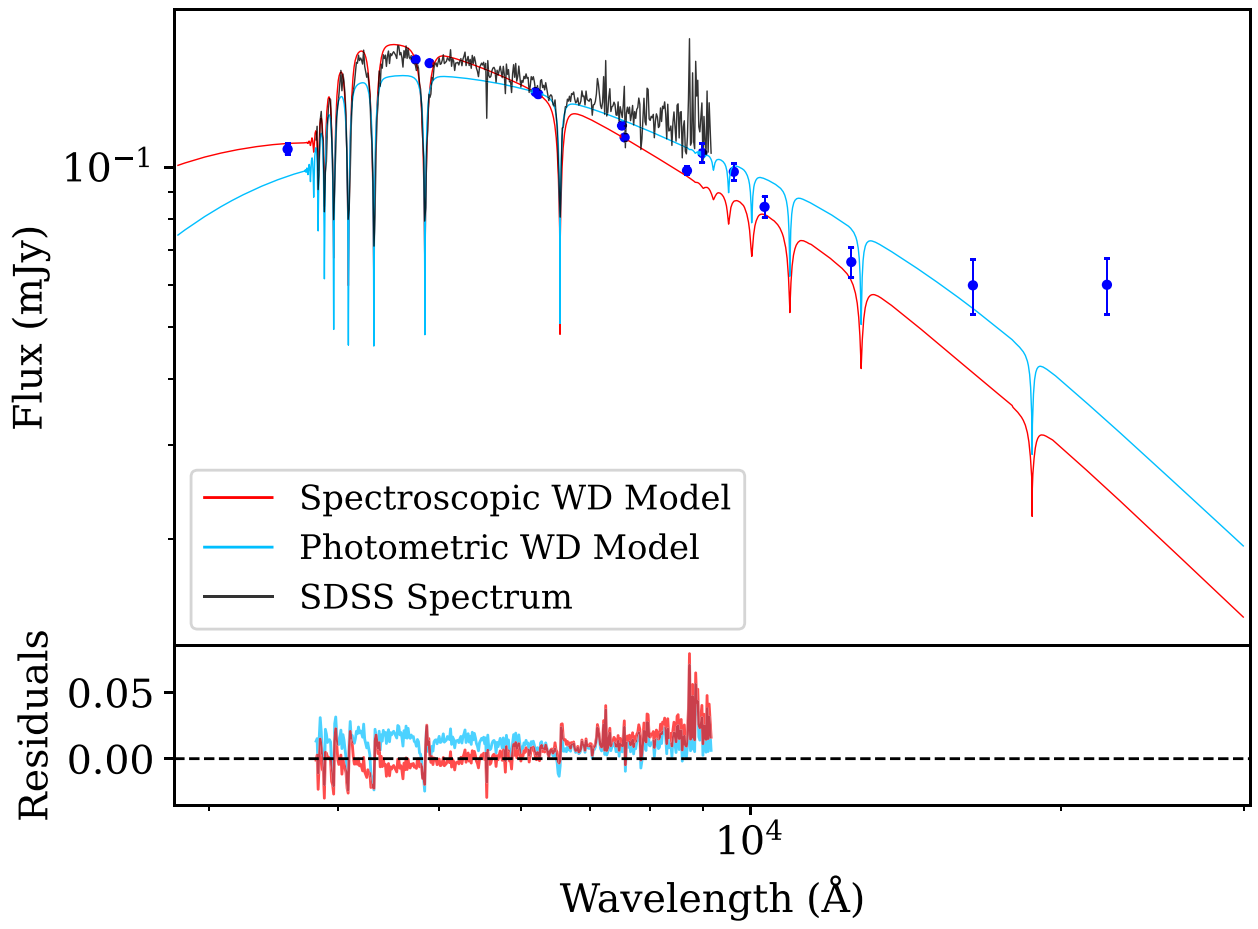

**Figure 2.** SDSS spectrum of SDSS J222551.65+001637.7A with dereddened SDSS, PAN-STARRS and UKIDSS photometry (blue). The Koester DA white dwarf model with $T_{\rm eff} = 9250$ K and $\log g = 7.75$, deriving from the photometric fits, is shown in red. The Koester DA white dwarf model with $T_{\rm eff} = 11000$ K and $\log g = 8.25$, deriving from the spectroscopic fits, is shown in light blue. The residuals between the SDSS spectrum and the white dwarf models are shown in the bottom panel.

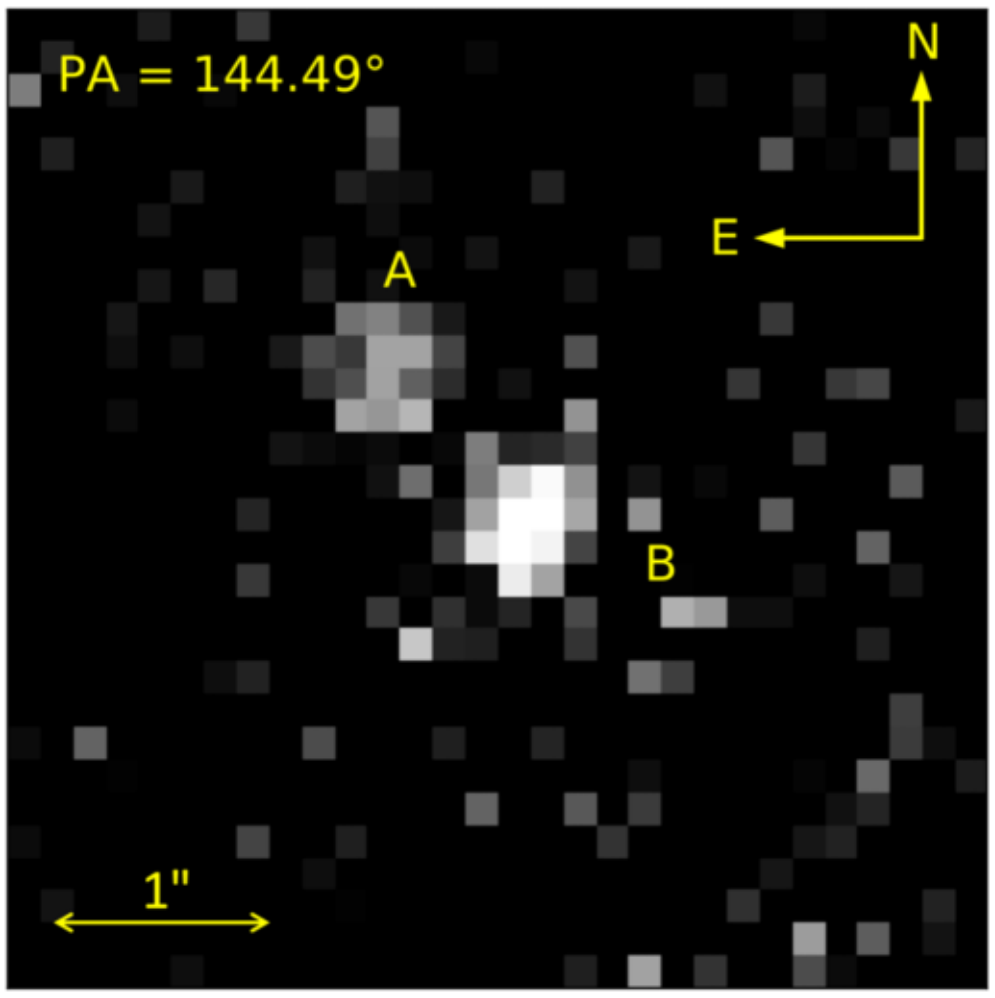

**Figure 3.** GNIRS *H*-band acquisition image of SDSS J222551.65+001637.7AB showing the resolved white dwarf and brown dwarf components of the binary. The position angle of the acquisition image is 144.49°. Both the white dwarf and the brown dwarf were in the slit during our observations.

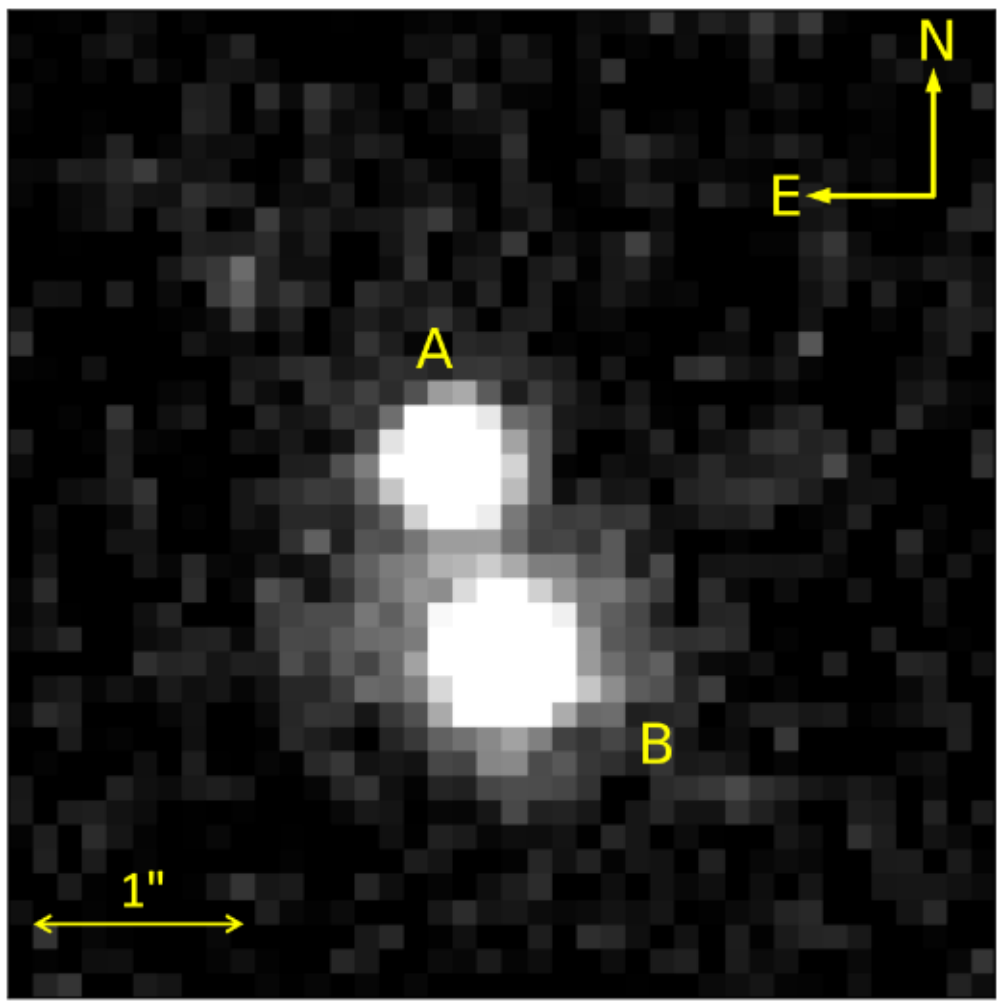

**Figure 4.** NIRI $K_s$-band image of SDSS J222551.65+001637.7AB showing we have clearly resolved the white and brown dwarf components of the binary. A is the white dwarf, and B is the brown dwarf.

## 3 OBSERVATIONS AND DATA REDUCTION

### 3.1 GNIRS spectroscopy

We observed SDSS J22255+0016AB using the cross-dispersed spectrograph GNIRS on Gemini North (Elias et al. 2006) on 2020 July 8th and July 10th UTC, as a part of programme GN-2020A-Q-322 (PI: John H. Debes). Spectra were taken using the short blue camera with the 32 l/mm grating and a slit width of 1.0 arcsec, giving a resolution of ($\lambda$ / $\Delta\lambda$) ∼ 500 across the entire wavelength range of 0.8–2.5 μm. We nodded the observations, taking 300 s exposures at each nod point, totalling 20 exposures. Arc lamp and flat-field calibration frames were taken immediately after the science observations. All exposures across three individual hours of observation were then combined during data reduction. Both the white dwarf and the brown dwarf were in the slit during our observations. The data were then reduced using a version of SPEXTOOL 4.1 (Cushing, Vacca & Rayner 2004) which has been adapted for use with GNIRS data (K. Allers, private communication). Two A0V standard stars, HIP115119 and HIP110963, were observed using the same settings with four frames taken for each hour of observation with an exposure time of 1.0 s. Two stars were observed to ensure that for each hour of science observation, the standard star observed was at a similar airmass. The telluric correction was performed during reduction using the XTELLCOR package (Vacca, Cushing & Rayner 2003) in SPEXTOOL using our observed standard stars.

The acquisition images from GNIRS revealed that the system is in fact spatially resolved (Fig. 3). The images were taken with 3 s exposures in the *H*-band using the same settings as the science observations.

### 3.2 NIRI imaging

We imaged SDSS J22255+0016AB on the 2021 June 13th UTC with NIRI in the $K_s$-band as part of programme GN-2021A-FT-207 (PI: Elena Manjavacas). We obtained 13 60 s exposures at airmass 1.09 with the f/6 camera providing a pixel scale of 117.1 mas pixel$^{-1}$.

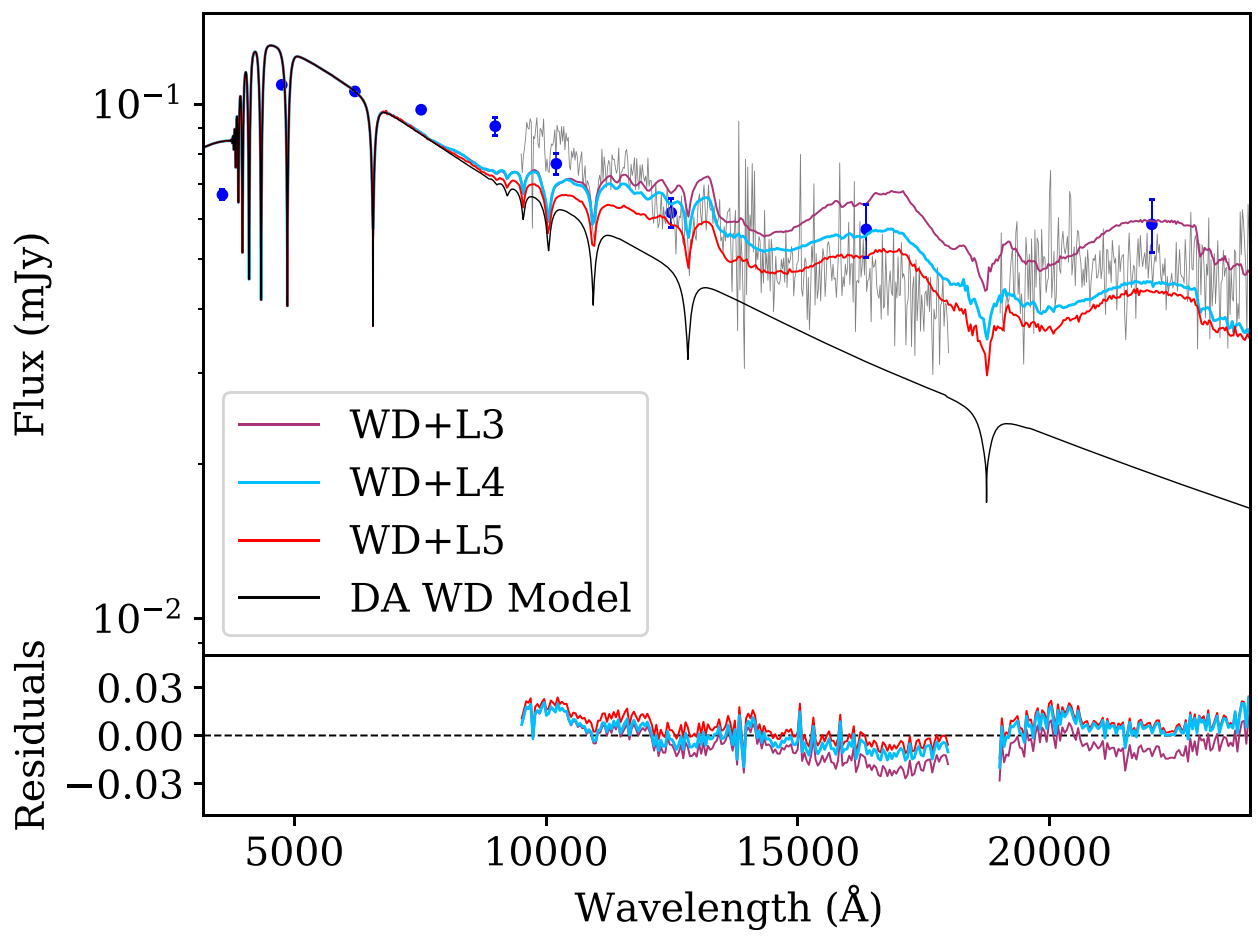

**Figure 5.** GNIRS spectrum of SDSS J222551.65+001637.7AB with SDSS and UKIDSS photometry (blue) and the Koester DA white dwarf model. The combined white dwarf+L3 template spectra, combined white dwarf+L4 template spectra and combined white dwarf+L5 template spectra are also shown. The residual spectra for L3–L5 spectral types are shown in the bottom panel. The GNIRS spectrum has been binned to a resolution of 1.4 Å to increase signal to noise, and the telluric line-dominated section between 18000 Å and 19000 Å has been masked for clarity.

We reduced the data using the DRAGONS software (Labrie et al. 2019) provided by the Gemini observatory, reducing using flat-field and dark frames provided as part of the calibration set and creating a bad pixel mask using 10 s dark frames. The 13 images were reduced and stacked using stars in the image as references. This is depicted in Fig 4. and shows that we have clearly resolved the white dwarf and brown dwarf components.

## 4 RESULTS

### 4.1 Spectral type

Although SDSS J22255+0016AB is resolved in the acquisition image, both of the components were within the slit during observations. To confirm the spectral type of the secondary, we created composite DA white dwarf + cool L-dwarf templates similar to those in Steele et al. (2011) and Casewell, Geier & Lodieu (2017). We used a hydrogen-rich DA white dwarf model (Koester 2010) with $T_{\rm eff} = 11\,000$ K and log $g = 8.25$ to best match the parameters of SDSS J22255+0016A. We took L-dwarf template spectra from the SpeX Prism Library (Burgasser 2014) for spectral types L3-L5, which are reported in Burgasser et al. (2010). We combined the white dwarf model with the L-dwarf template spectra by setting both to 10 pc and then combining them. To set the white dwarf model to 10 pc, we normalized using the *Gaia* eDR3 distance of $218^{+14}_{-13}$ pc and the broad-band photometry measurement in the SDSS $r$-band (Alam et al. 2015). To normalize the brown dwarf template spectra to 10 pc, we used the mean absolute $J$-band magnitudes for each spectral type from Dupuy & Liu (2012). Once combined, we normalized the composite white dwarf + L-dwarf models to the broad-band $r$-band photometry, and we normalized the GNIRS spectrum to the broad-band UKIDSS $J$-band photometry (Girven et al. 2011). We then compare the GNIRS spectrum to the composite white dwarf + L-dwarf models to determine the presence of an IR excess and the nature of the companion. Fig. 5 shows the GNIRS spectrum for SDSS J22255+0016AB alongside photometry measurements and our composite white dwarf + brown dwarf models. From this, we determine the spectral type of the brown dwarf as L4 ± 1.

### 4.2 Relative astrometry

From our stacked NIRI image (Fig. 4) , we measured relative astrometry for SDSS J22255+0016AB. We fitted an analytic point spread function model to each component, where the model used three concentric 2D Gaussians with different amplitudes, standard deviations, ellipticities, and angles for the ellipticities. This approach is based on previous work with adaptive optics imaging of low-mass binaries (e.g. Liu et al. 2006; Mann et al. 2019). We converted the pixel positions of the two components into sky coordinates using the WCS information in the FITS header. Given how well resolved SDSS J22255+0016AB is, the errors on its relative astrometry are dominated by the astrometric calibration of NIRI, with a fractional uncertainty of 0.23 percent in pixel scale and an uncertainty of 0.1° in parallactic angle (Mann et al. 2019). This results in a separation of 949.8 ± 2.2 mas and a parallactic angle of 194.6 ± 0.1° between the two components, measured as the position of B from A. Our binary fit also provides a measurement of the relative photometry of $K_{s,\rm B} - K_{s,\rm A} = -0.99 \pm 0.02$ mag. Here, we have defined the more massive white dwarf as the A component and its lower mass ultracool dwarf companion, which is brighter in the $K_s$-band, as the B component.

We do not derive relative astrometry from the GNIRS acquisition images, as it is not astrometrically well-calibrated. We note, however, that there does not appear to be any astrometric motion relative to the NIRI imaging. As the two images were taken 1 year apart, this is not surprising.

The two binary components are separated by 0.9498 ± 0.0022 arcsec and a parallactic angle of 194.6 ± 0.1° and have a magnitude difference of 0.990 mags in the $K_s$-band. This magnitude difference is consistent with the difference in absolute magnitudes of the two components as predicted by the models of Tremblay, Bergeron & Gianninas (2011) for DA white dwarfs and the absolute magnitude spectral type relations of Dupuy & Liu (2012) for an L4 brown dwarf. Using the *Gaia* eDR3 distance of $218^{+14}_{-13}$ pc and our separation of 0.9498 ± 0.0022 arcsec, we calculate the projected separation of SDSS J22255+0016AB as $207^{+13}_{-12}$ au.

## 5 DISCUSSION

We determined the spectral type of SDSS J22255+0016B as L4±1 by comparing template brown dwarf spectra to the GNIRS spectrum. This spectral type is consistent with the difference in magnitude for the components measured from the NIRI image. Although the $K$-band photometry and spectrum are brighter than that of the white dwarf + L4 combined model, it is consistent within the errors, which are dominated by the absolute magnitudes in Dupuy & Liu (2012). The offset between the models and the spectrum at 10000 Å is due to the SpeX template dwarf spectra not extending very far into optical wavelengths. Our spectral type is consistent with the L-dwarf companion proposed by Steele et al. (2011) based on the UKIDSS photometry. Our projected separation of $207^{+13}_{-12}$ au agrees with their prediction of a $<$ 350 au.

### 5.1 Age of the system

To determine the age of SDSS J22255+0016AB, we used WDWARFDATE, which estimates the age of a white dwarf, as well as its final mass and initial mass, from $T_{\rm eff}$ and log $g$ using a Bayesian framework (Kiman et al. 2022). The cooling age and mass of the white dwarf are determined from the evolutionary models of the Montreal White Dwarf Group (Bédard et al. 2020), and an initial-final mass relationship is used to calculate the initial mass of the white dwarf progenitor. The progenitor lifetime, also referred to as the main sequence (MS) age, is then determined using the MIST isochrones (Choi et al. 2016; Dotter 2016). The total lifetime of the white dwarf is calculated as the sum of the cooling age and the progenitor's MS age (Table 2). We utilized the initial-final mass relationship of Cummings et al. (2018) and assumed solar metallicity and $v/v_{\rm crit} = 0$ for the fit, where $v/v_{\rm crit}$ quantifies stellar rotation (Sun et al. 2021). The white dwarf mass of $0.66^{+0.11}_{-0.06}$ $M_\odot$ is within $1\sigma$ of the Anguiano et al. (2017) white dwarf mass of $0.72^{+0.10}_{-0.10}$ $M_\odot$. Our cooling age of the white dwarf gives the minimum age of the system as $0.58^{+0.17}_{-0.08}$ Gyr. We estimate the total age of the system as $1.97^{+4.41}_{-0.76}$ Gyr; however, this value is particularly sensitive to uncertainties in the choice of initial-final mass relationship and the MS age of the white dwarf progenitor.

**Table 2.** System parameters for SDSS J222551.65+001637.7AB derived using WDWARFDATE.

| Parameter | Value |
|---|---|
| Cooling Age (Gyr) | $0.58^{+0.17}_{-0.08}$ |
| Final Mass ($M_\odot$) | $0.66^{+0.11}_{-0.06}$ |
| Initial Mass ($M_\odot$) | $1.97^{+1.14}_{-0.76}$ |
| Main Sequence Age (Gyr) | $1.40^{+4.48}_{-0.98}$ |
| Total System Age (Gyr) | $1.97^{+4.41}_{-0.76}$ |

To further constrain the age of SDSS J22255+0016AB, we used the *Gaia* eDR3 proper motions and the radial velocity measured by Anguiano et al. (2017) to undertake a kinematic analysis. We calculate the UVW space velocities with respect to the local standard of rest as: $U = -9.52 \pm 7$ km s$^{-1}$, $V = 54.5 \pm 13$ km s$^{-1}$, $W = -71.8 \pm 15$ km s$^{-1}$. Here, U is positive towards the Galactic centre, V is positive in the direction of Galactic rotation, and W is positive towards the North Galactic Pole. Following the method of Bensby, Feltzing & Oey (2014) with their observed fractions of thick disc, thin disc, and halo populations in the solar neighbourhood, we determine the relative probabilities for SDSS J22255+0016AB belonging to each of these populations. We find that SDSS J22255+0016AB is 495 times more likely to belong to the thick disc than the thin disc and 461 times more likely to belong to the thick disc than the stellar halo. It is thus likely that SDSS J22255+0016AB is a member of the thick disc. The thick disc has an age of ~10 Gyr (Kilic et al. 2017), meaning that if SDSS J22255+0016AB is indeed a member of the thick disc, the total system age is likely closer to the upper uncertainty of the age we determine with WDWARFDATE. In their analysis of white dwarfs in the thin and thick discs, Raddi et al. (2022) find that the total age distribution of white dwarfs peaks at 2 Gyr, which may explain why the total age of SDSS J22255+0016AB is young for a thick disc object. Additionally, Torres et al. (2021) find that 13 percent of halo white dwarfs in *Gaia* DR2 are younger than expected compared to the average halo white dwarf age. This indicates the presence of younger white dwarfs in both disc and halo populations, of which SDSS J22255+0016A may be one; however, the origin of these younger objects is unclear. We note that older age is derived if we use the photometric parameters of the white dwarf; however, this has larger uncertainties, and the photometric parameters are less reliable due to their lack of reddening. Despite the large uncertainty in total age, which is dominated by uncertainties in the initial-final mass relationship, SDSS J22255+0016B is an important member in the small population of wide white dwarf–brown dwarf binaries.

**Table 3.** Absolute and apparent magnitudes for each component of SDSS J222551.65+001637.7AB.

| Star | Absolute $K_s$ Magnitude | Apparent $K_s$ Magnitude |
|---|---|---|
| White Dwarf | $12.26 \pm 0.20$ | $18.96 \pm 0.15$ |
| Brown Dwarf | $11.27 \pm 0.18$ | $17.97 \pm 0.13$ |

As a member of the thick disc, it is unlikely that SDSS J22255+0016AB is extremely metal-poor in comparison to objects residing in the stellar halo. There is no evidence of photospheric metal pollution in the SDSS optical spectrum of the white dwarf that would indicate accretion from a tidally disrupted asteroid or another companion (Zuckerman et al. 2003; Debes 2006). SDSS J22255+0016A has a low effective temperature, and if it were polluted, absorption features would be easily detectable in the optical spectrum. Since we do not detect any pollution, it is thus likely that SDSS J22255+0016AB has no other companions. However, we note that the Ca II line can appear weak in white dwarf spectra, and a high resolution echelle spectrum of the white dwarf would be required to place definitive limits on any potential pollution (Zuckerman et al. 2003).

### 5.2 SDSS J222551.65+001637.7B

As discussed in Section 4.2, our observed absolute magnitude difference between the white dwarf and the brown dwarf is consistent with predictions from theoretical models. Using this magnitude difference and taking our observed UKIDSS *K*-band magnitude of $17.6 \pm 0.13$ as a proxy for the observed $K_s$-band magnitude, we calculate the apparent magnitudes of both the white dwarf and the brown dwarf. Using the *Gaia* distance of $218^{+14}_{-13}$ pc, we calculate the absolute magnitudes in the $K_s$-band for both the white dwarf and the brown dwarf. These magnitudes are presented in Table 3. We find that absolute $K_s$-band magnitude of the brown dwarf is $11.27 \pm 0.18$. This is consistent with the mean absolute $K_s$-band magnitude of $11.55 \pm 0.28$, which Dupuy & Liu (2012) report for L4 companions. Our absolute $K_s$-band magnitude for the white dwarf is $12.26 \pm 0.18$. This is consistent with that predicted by synthetic photometry calculated from the Tremblay et al. (2011) white dwarf models (Holberg & Bergeron 2006; Kowalski & Saumon 2006). [1]

The estimated L4 companion spectral type provides a consistent theoretical and observed absolute magnitude, indicating that SDSS J22255+0016B is indeed an L4 ± 1 companion. We estimate the effective temperature of SDSS J22255+0016B as $T_{\rm eff} = 1800^{+70}_{-60}$ K for our spectral type of L4 ± 1 as this is the mean effective temperature of an L4 dwarf determined from the analysis of M, L, and T dwarfs performed by Vrba et al. (2004). We then compare our estimated effective temperature and our $K_s$-band magnitude for an L4 spectral type to the Sonora–Bobcat models, assuming solar metallicity (Marley et al. 2021). From these models, a brown dwarf with $T_{\rm eff} = 1800$ K and $K_s = 11.27$ would have a mass of 43.88 $M_{\rm Jup}$ and a radius of 0.1071 $R_\odot$. This mass estimate is consistent with the mass of $47 \pm 3\,M_{\rm Jup}$ determined by Steele et al. (2011) using the Lyon group models.

With our $K_s$-band magnitude, we use the relations of Dupuy & Liu (2017) to calculate the bolometric luminosity of

[1] http://www.astro.umontreal.ca/~bergeron/CoolingModels

SDSS J22255+0016B as log ($L_{\rm bol}/L_\odot$) = −3.92 ± 0.11. We also utilize their Lyon $T_{\rm eff}$ relation to improve our temperature estimate to $T_{\rm eff}$ = 1817 ± 90 K. We compare our bolometric luminosity and effective temperature of the brown dwarf with the Sonora–Bobcat models, assuming solar metallicity (Marley et al. 2021). From these models, a brown dwarf with log ($L_{\rm bol}/L_\odot$) = −3.92 ± 0.11 and $T_{\rm eff}$ = 1817 ± 90 K would have a mass of 25–53 $M_{\rm Jup}$ and a radius of 0.101–0.128 $R_\odot$. We find that the most appropriate model for our bolometric luminosity and effective temperature provides an age estimate and a $K_s$-band magnitude that are consistent with our results.

### 5.3 Evolution of the system

During the evolution of the MS progenitor of SDSS J22255+0016A, the orbital separation would have increased by a maximum factor of $M_{\rm MS}/M_{\rm WD}$ = 2.99 (Burleigh, Clarke & Hodgkin 2002). The initial projected separation would therefore have been >69 au, confirming that this is not a post-common envelope binary. Burleigh et al. (2002) state that white dwarfs will retain their planetary companions if the initial separation from the MS progenitor star is >5 au, as is the case for SDSS J22255+0016AB.

Since the initial separation of SDSS J22255+0016AB is too wide to be a post-common envelope system, it will have evolved differently to close white dwarf–brown dwarf binaries (e.g. Maxted et al. 2006; Casewell et al. 2018). The two components will have evolved separately, and the brown dwarf will not have truncated the white dwarf's evolution. However, the brown dwarf may have been affected by stellar winds from the primary, with the angular momentum lost by the white dwarf causing the separation to increase (Schrøder et al. 2021). During the evolution of the MS progenitor of the white dwarf, the star undergoes a phase of evolution on the Asymptotic Giant Branch (AGB) before reaching its end stage as a white dwarf (Iben & Renzini 1983). In the AGB phase, the mass-loss increases until the envelope is fully ejected, which causes stellar winds that can affect the substellar companion. Mayer et al. (2014) found dust-enriched winds of $v_w$ = 5–20 km s$^{-1}$ and Höfner & Olofsson (2018) report outflowing winds between $v_w$ = 3–30 km s$^{-1}$, affecting companions at separations on the order of ∼100 au. It is possible for the presence of the companion to shape these winds, morphing spherical AGB stars into non-spherical planetary nebulae, but at these wide separations ($\gtrsim$ 50 au), this is unlikely to alter the white dwarf progenitor's evolution (Decin et al. 2020).

### 5.4 Orbit

Using the white dwarf mass, the brown dwarf mass we estimate from the Sonora–Bobcat models, and our projected orbital separation of $207^{+13}_{-12}$ au, we calculate the likely orbital period of SDSS J22255+0016AB as $P$ = 3560 ± 383 yr. This is a minimum period assuming a circular orbit; however, many brown dwarfs are in eccentric orbits. Ma & Ge (2014) report that the eccentricity distribution of brown dwarfs changes at a threshold mass of 42.5 $M_{\rm Jup}$, with brown dwarfs below this mass having eccentricities similar to massive planets and brown dwarfs above this mass having eccentricities consistent with binaries. This suggests two distinct formation mechanisms for brown dwarfs: protoplanetary discs and stellar binary-like formation. SDSS J22255+0016B resides near this mass boundary, and further investigations such as continuous monitoring to calculate its dynamical mass and observations to obtain an uncontaminated spectrum of the brown dwarf and a C/O ratio measurement, would enable us to determine its formation mechanism. The mass ratio of SDSS J22255+0016AB is $q = M_{\rm BD}/M_{\rm MS}$ = 0.012–0.048. Bowler, Blunt & Nielsen (2020) state that for binary mass ratios >0.01 stellar binary-like formation is favoured, which also indicates a higher eccentricity than systems in which the brown dwarf formed via planet-like formation.

Fig. 6 depicts the currently known white dwarf–brown dwarf binaries as well as directly imaged brown dwarfs and exoplanets around MS stars. The white dwarf–brown dwarf binaries are colour coded according to the spectral type of the brown dwarf. The outlined star represents SDSS J22255+0016AB in its evolved form as it is now. The outlined pentagon represents SDSS J22255+0016AB whilst the white dwarf progenitor was still on the MS, with $M_{\rm WD}$ = 1.97 $M_\odot$ and a binary separation of 69 au. We identify four directly imaged brown dwarfs around MS stars that are similar to SDSS J22255+0016AB before the primary star evolved into a white dwarf. These systems are outlined in black and are, top to bottom, HD 19467AB, HD 33632AB, HR 3549AB, and GJ 758AB. A comparison of these four systems with the progenitor of SDSS J22255+0016AB is made in Table 4. These four systems all have similar mass ratios, separations and companion masses to SDSS J22255+0016AB before the white dwarf progenitor evolved into the white dwarf, increasing the separation of the brown dwarf as it evolved. It is therefore likely that SDSS J22255+0016AB formed via a similar mechanism to these binaries, which all formed in stellar-like or stellar binary-like mechanisms (Vigan et al. 2016; Currie et al. 2020; Maire et al. 2020). Additionally, when the MS stars in HD 19467AB, HD 33632AB, HR 3549AB, and GJ 758AB evolve into a white dwarf, their evolved forms will resemble SDSS J22255+0016AB. In particular, HR 3549AB will be most comparable to SDSS J22255+0016AB once it has evolved. HR 3549A has a white dwarf mass within 1$\sigma$ of the mass of SDSS J22255+0016A and a brown dwarf mass within the mass range of SDSS J22255+0016B. Of the four objects highlighted here, HR 3549AB has a separation most akin to the estimated initial separation of SDSS J222551.65+001637.7. Furthermore, HR 3549AB is younger than SDSS J22255+0016AB, and the brown dwarf has an earlier spectral type, meaning it could conceivably evolve into an extremely similar system over time.

With a projected separation of $207^{+13}_{-12}$ au between the two components, SDSS J22255+0016AB is the third closest separated spatially resolved wide white dwarf–brown dwarf binary after GD 165AB (Becklin & Zuckerman 1988) and PHL 5038AB (Steele et al. 2009). These 3 systems comprise a subset of wide, but not ultra-wide, white dwarf–brown dwarf binaries which are spatially resolved, as opposed to the other 5 ultra-wide, comoving, resolved systems currently known. Table 5 details the parameters of these 3 systems. SDSS J22255+0016AB is most similar to GD 165AB, with comparable white dwarf masses, effective temperatures and surface gravities, as well as total ages. However, GD 165B has a higher mass and smaller physical separation than SDSS J22255+0016B. Although the brown dwarf in PHL 5038AB is a later spectral type, and its white dwarf primary is cooler than SDSS J22255+0016A, these binaries are still akin to each other, with separations on the order of 100 au, and white dwarf masses and surface gravities within 1$\sigma$ of each other. The dominant factor influencing the evolution of these binaries is the white dwarf mass and the separation, since at wide separations the brown dwarf is not massive enough to affect the evolution of the binary. It is thus likely that these three resolved systems all evolved in the same manner.

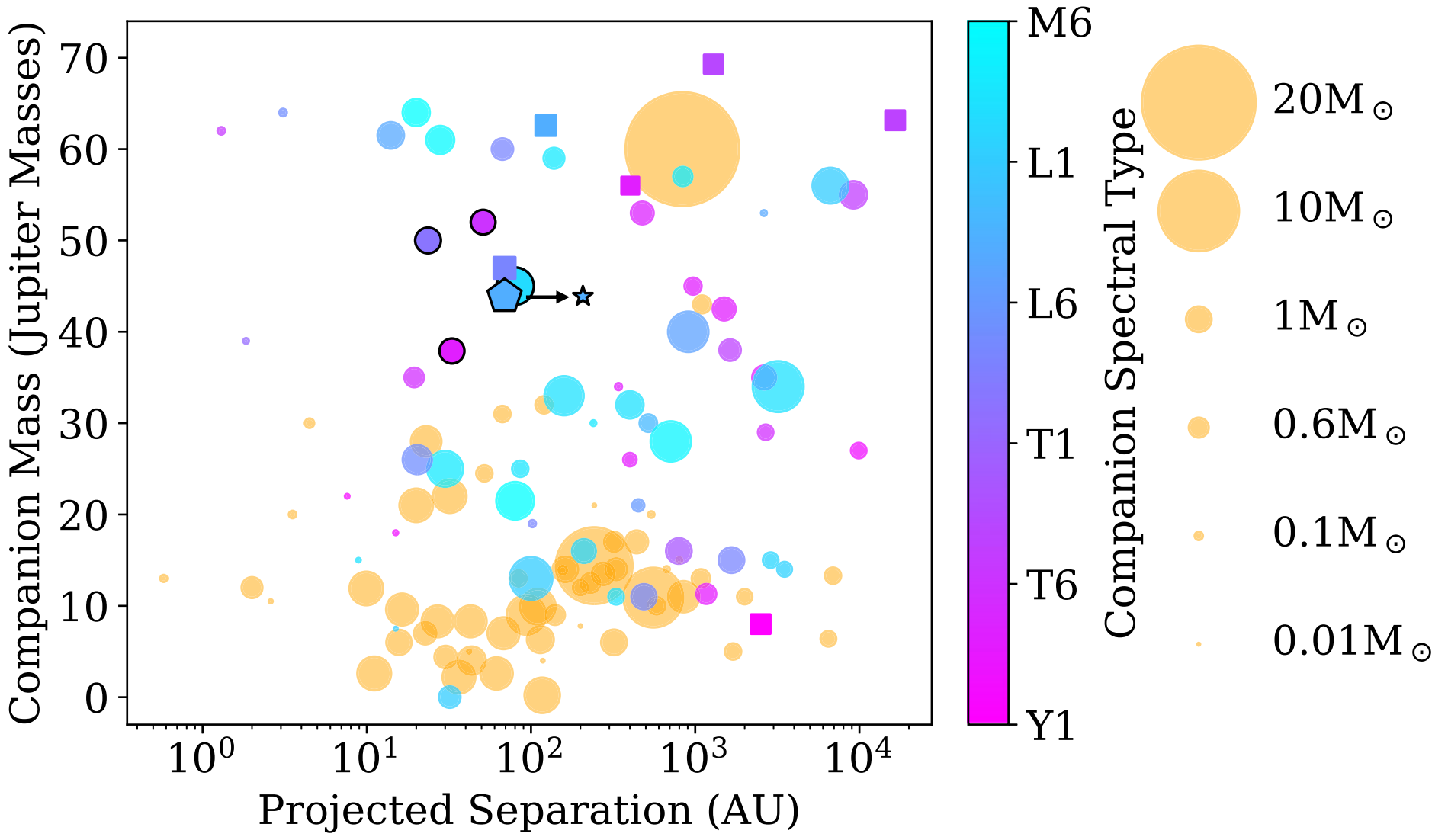

**Figure 6.** Distribution of known white dwarf–brown dwarf binaries alongside directly imaged exoplanets and binaries around main sequence stars. Exoplanets are in orange and brown dwarfs are colour coded by their spectral type. Directly imaged objects are represented by circles and the white dwarf–brown dwarf binaries are represented by squares. Point size is proportional to the mass of the primary star. The star represents SDSS J222551.65+001637.7AB at present. The pentagon represents SDSS J222551.65+001637.7AB whilst the white dwarf progenitor was still on the main sequence. The outlined circles are the four systems most similar to the progenitor of SDSS J222551.65+001637.7AB: HD 19467AB, HD 33632AB, HR 3549AB, and GJ 758AB.

**Table 4.** Comparison of the four directly imaged main sequence-brown dwarf binaries most similar to SDSS J222551.65+001637.7AB: HD 19467, HD 33632, HR 3549, and GJ 758. SDSS J222551.65+001637.7AB is reported as it was when the white dwarf progenitor was still on the main sequence.

| Binary | $M_{\rm MS}$ ($M_\odot$) | Age (Gyr) | BD Spectral Type | $M_{\rm BD}$ ($M_{\rm Jup}$) | Separation (au) | Ref |
|---|---|---|---|---|---|---|
| HD 19467AB | $0.953 \pm 0.022$ | $5.4^{+1.9}_{-1.3}$ | T5.5 | $65.4^{+5.9}_{-4.6}$ | $51.1 \pm 0.1$ | 3, 4, 5, 6 |
| HD 33632AB | $1.1 \pm 0.1$ | $1.7 \pm 0.4$ | L9.5 | $50.0^{+5.6}_{-5.0}$ | $23.6^{+3.2}_{-4.5}$ | 4, 7 |
| HR 3549AB | $2.3 \pm 0.2$ | $0.10 - 0.15$ | M9.5 | $45 \pm 5$ | $80.0 \pm 2.0$ | 1, 2 |
| GJ 758AB | $0.96 \pm 0.3$ | $8.3^{+2.7}_{-2.1}$ | T8 | $38.0 \pm 0.8$ | $33.0 \pm 6.0$ | 4, 8, 9, 10 |
| SDSS J222551.65+001637.7AB | $1.97^{+1.14}_{-0.76}$ | $1.40^{+4.48}_{-0.98}$ | L4 | 25–53 | $69 \pm 5$ | This Work |

*Note.* References are 1: Mawet et al. (2015); 2: Mesa et al. (2016); 3: Maire et al. (2020); 4: Brandt et al. (2021); 5: Crepp et al. (2015); 6: Jensen-Clem et al. (2016); 7: Currie et al. (2020); 8: Takeda (2007); 9: Vigan et al. (2016); 10: Brandt, Dupuy & Bowler (2019).

**Table 5.** Comparison of the three closest separated resolved white dwarf–brown dwarf binary systems, GD 165AB, PHL 5038AB, and SDSS J222551.65+001637.7AB.

| Binary | $M_{\rm WD}$ ($M_\odot$) | $T_{\rm eff}$ (K) | log $g$ | $M_{\rm BD}$ ($M_{\rm Jup}$) | Spectral Type | Separation (au) | Age (Gyr) | Ref |
|---|---|---|---|---|---|---|---|---|
| GD 165AB | $0.64 \pm 0.02$ | $12130 \pm 450$ | $8.052 \pm 0.035$ | $62.58 \pm 15.57$ | L4 | $123 \pm 12$ | 1.2–5.5 | 1, 2, 3, 4, 5 |
| PHL 5038AB | $0.72 \pm 0.15$ | $8000 \pm 100$ | $8.2 \pm 0.1$ | 60 | L8 | $69 \pm 1$ | 1.9–2.7 | 6 |
| SDSS J222551.65+001637.7AB | $0.66^{+0.11}_{-0.06}$ | $10926 \pm 246$ | $8.214 \pm 0.168$ | 25–53 | L4 | $207^{+13}_{-12}$ | 1.2–6.4 | 7, This Work |

*Note.* References are 1: Giammichele et al. (2016); 2: Filippazzo et al. (2015); 3: Becklin & Zuckerman (1988); 4: Kirkpatrick et al. (1993); 5: Kirkpatrick et al. (1999); 6: Steele et al. (2009); 7: Anguiano et al. (2017).

## 6 CONCLUSIONS

We confirm SDSS J222551.65+001637.7AB as a wide, comoving white dwarf–brown dwarf binary, which has now become resolved. Alongside the photometry measurements, the near-IR spectrum taken by GNIRS shows an IR excess that indicates a brown dwarf companion of spectral type L4 ± 1. We determine the absolute $K_s$-band magnitude of the brown dwarf as $11.27 \pm 0.18$, which is consistent with an L4 ± 1 spectral type. We calculate the white dwarf mass as $0.66^{+0.11}_{-0.06}$ $M_\odot$ and the total system age as $1.97^{+4.41}_{-0.76}$ Gyr. We use the Sonora–Bobcat evolutionary models to estimate the mass of the companion as 25–53 $M_{\rm Jup}$ and its radius as 0.101–0.128 $R_\odot$, confirming that it is a brown dwarf. The white dwarf shows no metal-line pollution that would indicate the presence of another companion. The acquisition image from the GNIRS spectrum and subsequent NIRI imaging confirm that SDSS J222551.65+001637.7AB is spatially resolved with an angular separation of $0.9498 \pm 0.0022$ arcsec, which corresponds to a projected separation of $207^{+13}_{-12}$ au at the *Gaia* eDR3 distance of $218^{+14}_{-13}$ pc. We calculate UVW space velocities to demonstrate that this system is likely a member of the thick disc. We estimate the minimum orbital period of this binary as $P = 3560 \pm 383$ yr. Due to the wide separation, it is unlikely that the brown dwarf companion altered the primary progenitor's evolution. This system is only the 8th confirmed wide comoving white dwarf–brown dwarf binary and constitutes the third closest separated resolved system after GD 165AB (Becklin & Zuckerman 1988) and PHL 5038AB (Steele et al. 2009).

## ACKNOWLEDGEMENTS

The authors would like to thank Kathleen Labrie of Noirlab for her help in reducing the data using the Dragons pipeline. JRF acknowledges support of a University of Leicester College of Science and Engineering PhD studentship. SLC acknowledges the support of a Science and Technology Facilities Council Ernest Rutherford Fellowship (ST/R003726/1). ECM acknowledges the support of the Heising Simons Foundation 51 Pegasi b Fellowship (#21-0684). This work is based on observations obtained at the international Gemini Observatory, a program of the National Science Foundation's National Optical-Infared Astronomy Research Laboratory, which is managed by the Association of Universities for Research in Astronomy (AURA) under a cooperative agreement with the National Science Foundation on behalf of the Gemini Observatory partnership: the National Science Foundation (United States), National Research Council (Canada), Agencia Nacional de Investigación y Desarrollo (Chile), Ministerio de Ciencia, Tecnología e Innovación (Argentina), Misitério da Ciência, Tecnologia, Inovações e Comunicações (Brazil), and Korea Astronomy and Space Science Institute (Republic of Korea). This work was enabled by observations made from the Gemini North telescope, located within the Maunakea Science Reserve and adjacent to the summit of Maunakea. We are grateful for the privilege of observing the Universe from a place that is unique in both its astronomical quality and its cultural significance. This research has made use of data obtained from or tools provided by the portal exoplanet.eu of The Extrasolar Planets Encyclopaedia. We thank the anonymous reviewer for their helpful comments which improved the manuscript.

## DATA AVAILABILITY

All data in this paper are publicly available in the Gemini archive.